\def\F        {{$^{19}$F \/}}

\def\CaF      {{CaF$_2$ \/}}
\def\etal     {{\it et al. \/}}

\newcommand{\mr}[1]{\mathrm{#1}}
\newcommand{\unit}[1]{\,\mathrm{#1}}
\newcommand{\um}{\,\mu{\rm m}}

\newcommand{\kT}{k_{\rm B}T}
\newcommand{\kB}{k_{\rm B}}

\newcommand{\fci}{f_{i}}
\newcommand{\kci}{k_{i}}
\newcommand{\Qci}{Q_{i}}
\newcommand{\tm}{\tau_\mr{m}}
\newcommand{\itm}{\tau_\mr{m}^{-1}}
\newcommand{\wrf}{\omega_\mr{rf}}
\newcommand{\weff}{\omega_\mr{eff}}

\newcommand{\wo}{\omega_0}
\newcommand{\wone}{\omega_1}

\newcommand{\Dw}{\Delta\omega}

\newcommand{\Tr}{T_{1\rho}}

\newcommand{\dxBu}{\frac{\partial {\bm B}}{\partial x}}
\newcommand{\dxBzu}{\frac{\partial B_z}{\partial x}}

\newcommand{\SBz}{S_{Bz}(\wone)}
\newcommand{\Geff}{G_\mr{eff}}
\newcommand{\Btip}{{\bm B}_\mr{tip}}

\newcommand{\Bext}{B_\mr{ext}}

\newcommand{\vecr}{{\bm r}}
\newcommand{\xrms}{x_\mr{rms}}
\newcommand{\Tmode}{T_\mr{mode}}

\documentclass[prl,aps,twocolumn,showpacs]{revtex4}

\usepackage{graphicx}
\usepackage{bm}
\usepackage{amsmath}
\usepackage{amssymb}
\usepackage[colorlinks=true, pdfstartview=FitV, linkcolor=blue, citecolor=blue, urlcolor=blue]{hyperref} 

\begin{document}

\global\emergencystretch = .1\hsize 

\title{Nuclear spin relaxation induced by a mechanical resonator}

\author{C. L. Degen$^{1}$}
  \email{degenc@gmail.com} 
\author{M. Poggio$^{1,2}$, H. J. Mamin$^1$, and D. Rugar$^1$}
  \affiliation{
   $^1$IBM Research Division, Almaden Research Center, 650 Harry Road, San Jose CA, 95120, USA. \\
   $^2$Center for Probing the Nanoscale, Stanford University, 476 Lomita Hall, Stanford CA, 94305, USA.}
\date{\today}

\begin{abstract}
We report on measurements of the spin lifetime of nuclear spins strongly coupled to a micromechanical cantilever as used in magnetic resonance force microscopy. We find that the rotating-frame correlation time of the statistical nuclear polarization is set by the magneto-mechanical noise originating from the thermal motion of the cantilever. Evidence is based on the effect of three parameters: (1) the magnetic field gradient (the coupling strength), (2) the Rabi frequency of the spins (the transition energy), and (3) the temperature of the low-frequency mechanical modes.
Experimental results are compared to relaxation rates calculated from the spectral density of the magneto-mechanical noise.
\end{abstract}


\maketitle

Sensitive detection of nuclear spin signals requires a sensor that can couple strongly to the weak nuclear magnetic moment.
For inductively detected nuclear magnetic resonance (NMR), where a resonant coil picks up the rf signal of the precessing spins,
this is achieved by scaling down the coil size so as to improve the current-per-flux ratio \cite{hoult76}.
Optimized inductive detectors are currently able to observe ensembles containing roughly $10^{12}$ proton spins,
equivalent to about $10^8$ net magnetic moments \cite{seeber01}.
Stronger couplings enabling higher spin sensitivity have been shown with magnetic force sensors,
such as used in magnetic resonance force microscopy (MRFM)  \cite{sidles95,rugar04}, which recently
measured the net moment of about $10^3$ nuclei \cite{mamin07}.

In order to detect spins with still higher sensitivity, the detector must be coupled even more tightly to the nuclear magnetic moment.
For a mechanical detector sensitive to forces in the $x$-direction, this is achieved by increasing the magnetic field gradient $\dxBu$,
which generates the magnetic force $F_x = {\bm \mu}\cdot\dxBu$,
where ${\bm \mu}$ is the total magnetic moment of the spin ensemble.
Present MRFM technology employs gradients up to $10^6\unit{T/m}$ \cite{mamin07}.
If the gradient can be successfully pushed into the range of $10^8\unit{T/m}$,
force detection of single nuclear spins may become feasible.

The strong interaction between spins and sensor, however, also increases the ``back action'' of the sensor on the spins and makes
them more susceptible to detector noise. Here we are concerned with magnetic noise generated by the thermally vibrating
cantilever, but similar effects can be expected in any real-world detector, as for example in an inductively-coupled rf circuit.
The interaction between a mechanical resonator and spins has been the subject of a number of theoretical studies,
and is predicted to lead to a host of intriguing effects. These range from shortening of spin lifetimes \cite{mozyrsky03,berman03},
to spin alignment by specific mechanical modes either at the Larmor frequency or in the rotating frame \cite{magusin00,butler05},
to resonant amplification of mechanical oscillations \cite{bargatin03}.

In this letter, we report direct experimental evidence for accelerated nuclear spin relaxation induced by a single, low-frequency mechanical mode.
The observed relaxation has its origin in the random magnetic field ${\bm B}(t) = \dxBu x(t)$
created by the thermal (Brownian) motion $x(t)$ of the cantilever in a large field gradient.
The nuclear spin and mechanical oscillator degrees of freedom are well decoupled in strong static fields,
since the spin precession frequency is orders of magnitude higher than typical cantilever frequencies.
However, when an on-resonance transverse rf magnetic field $B_1$ is present,
kHz frequency noise that overlaps with the Rabi frequency of the spin will induce spin relaxation.
The rate of nuclear spin transitions may be described by the rotating-frame relaxation time $\Tr$,
\begin{equation}
\Tr^{-1} \approx \frac{\gamma^2}{2} \SBz,
\label{eq_tr}
\end{equation}
where $\wone = \gamma B_1$ is the Rabi frequency of the spin, $\gamma$ is the gyromagnetic ratio,
and $\SBz$ is the (double-sided) power spectral density of the $z$-component of ${\bm B}(t)$ \cite{slichter_specdens,mozyrsky03}.

\begin{figure}
      \begin{center}
      \includegraphics[width=0.45\textwidth]{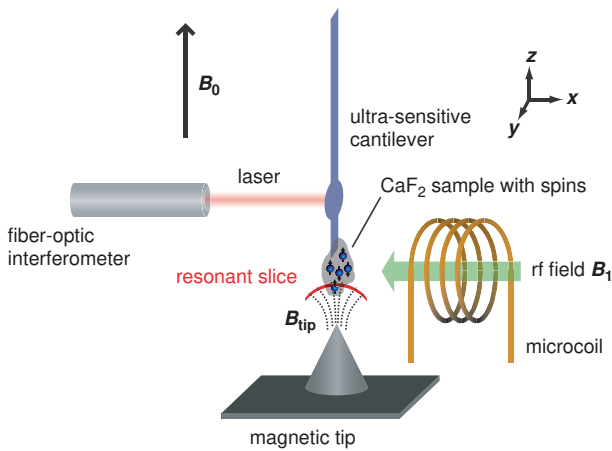}
      \end{center}
      \caption{(a) Experimental arrangement (setup {\bf A}). A \CaF single crystal containing \F nuclei
      is attached to the end of the cantilever and placed in a fixed position $\sim100\unit{nm}$ above a nanoscale FeCo magnetic tip.
      An rf magnetic field induces magnetic resonance in a thin ``resonant slice'' of spins where the Larmor resonance condition is fulfilled.
      Raising or lowering the external field $\Bext$ shifts the resonant slice position up or down,
      allowing spins further from or closer to the tip to be selectively addressed, respectively.
      }
      \label{fig_schematics}
\end{figure}

For a transversely oscillating cantilever as shown in Fig. \ref{fig_schematics},
$\SBz$ is related to the spectrum of cantilever tip motion by $\SBz = \left(\dxBzu\right)^2 S_x(\wone)$,
and will --- if $S_x(\wone)$ has thermal origin --- be proportional to $\kT$.
Hence, in a strong coupling regime where the spin lifetime is dominated by noise from the mechanical resonator,
we expect an explicit dependence on the field gradient, the Rabi frequency, the temperature,
and the mechanical mode spectrum.

We find evidence for mechanically induced spin relaxation while measuring nuclear spin correlation times for small ensembles
of statistically polarized \F spins in a \CaF single-crystal sample \cite{bloch46,degen07}.
The setup for these experiments, shown in Fig. \ref{fig_schematics}(a), combines three components: 
(1) an ultra-sensitive cantilever, (2) a nanoscale ferromagnetic tip, and (3) a micron-scale rf circuit for rf field generation.
Experiments are carried out with two different arrangements: setup {\bf A} uses a 90-$\um$-long single-crystal Si cantilever \cite{chui03}
together with a FeCo thin-film conical tip and an external rf microcoil \cite{mamin07}.
Setup {\bf B} employs a slightly longer cantilever ($120\unit{\um}$)
combined with a FeCo cylindrical pillar integrated onto a lithographically patterned rf microwire \cite{poggio07}.
The \CaF samples, a few $\um^3$ in size, are glued to the end of the cantilevers.

For spin detection we rely on the ``rf frequency sweep'' method to drive adiabatic spin inversions,
thereby modulating the $z$-component of the nuclear polarization at the fundamental cantilever frequency \cite{madsen04,poggio07}.
Here, the rf center frequency $\wrf/2\pi$ is $114\unit{MHz}$ with peak frequency deviation of the rf sweep $\Dw/2\pi$ in the range of $400$ to $1000\unit{kHz}$.
We measure the correlation time $\tm$ of the cantilever tip oscillation amplitude,
which reflects the correlation time of the nuclear polarization \cite{degen07}.
$\tm$ can be determined in several ways, for example by calculating the autocorrelation function \cite{degen07},
by measuring the linewidth of the associated power spectral density, or by using a bank of filters of different noise bandwidths \cite{rugar04}.
$\tm$ is very closely related to the rotating frame relaxation time $\Tr$, as will be discussed below.

\begin{figure}[t]
      \begin{center}
      \includegraphics[width=0.45\textwidth]{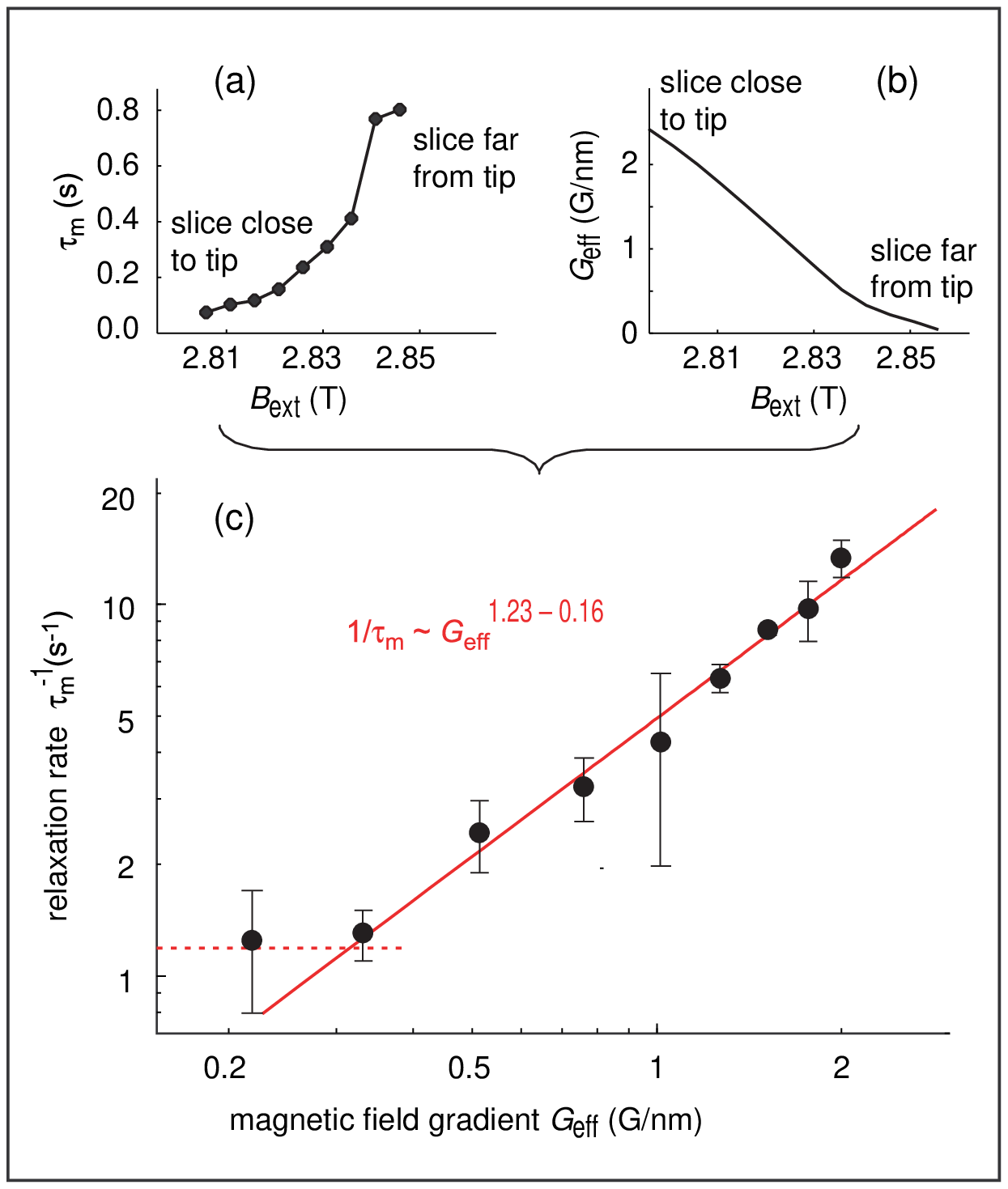}
      \end{center}
      \caption{
      (a) Spin correlation time $\tm$ as a function of external field $\Bext$.
      Data points (dots) represent experimental values, measured with setup {\bf A}.
      Spins are resonant at a total field of $|\Bext\hat{\bm z}+\Btip|=2.85\unit{T}$.
      (b) Effective field gradient $\Geff(\Bext)$ as a function of $\Bext$ calculated from tip model.
      (c) Relaxation rate $\itm$ as a function of field gradient $\Geff$ obtained by combining (a) and (b).
      Solid line is a best fit with the lowest point excluded. Dashed line is a guide to the eye.
      }
      \label{fig_gradient}
\end{figure}

In a first set of experiments we observe that spin relaxation depends strongly on the distance
between spins and magnetic tip. We attribute this observation
to an associated variation of the magnetic field gradient.
Figure \ref{fig_gradient}(a) plots the relaxation rate $\itm$ as a function of applied external field $\Bext$.
Since the external field sets the region of space where the resonance condition is met ---
indicated by the ``resonant slice'' in Fig. \ref{fig_schematics} ---
changing $\Bext$ is equivalent to changing the region in the sample where we probe the spins.
Spins located closer to the tip require less external field to satisfy the Larmor condition,
$\gamma |\Bext \hat{\bm z} +\Btip| = \wrf$.
For these spins, the resonance appears on the low end of the $\Bext$ scan [Fig. \ref{fig_gradient}(a)].
Likewise, spins far away from the tip experience the smallest $\Btip$ and require the highest $\Bext$.

In order to relate $\itm$ to the magnetic field gradient, we first calculate the tip field $\Btip(\vecr)$ as a function of position $\vecr$.
A good model for the tip can be obtained by inferring the tip size from a scanning electron micrograph,
and combining it with MRFM data for the tip moment \cite{mamin07}.
To estimate the effective lateral field gradient $\Geff(\Bext) \propto \dxBzu$ for each resonant slice,
we then derive the gradient from $\Btip(\vecr)$ and average it over the resonant slice volume \cite{Geff}.
As a result, we can relate the gradient to the external field and use this knowledge to plot $\itm$ as a function of $\Geff$,
shown in Fig. \ref{fig_gradient}(b) and (c).

We find that spin relaxation increases with the gradient as $\itm\propto \Geff^{1.23\pm0.16}$.
Since the gradient is the main parameter describing the coupling between spins and oscillator,
a stronger gradient is equivalent to a stronger coupling --- hence the same mechanical noise creates
more magnetic noise. While gradient-induced spin relaxation is the dominant relaxation mechanism
in Fig. \ref{fig_gradient}, other (intrinsic) processes will eventually dominate in the limit of small gradient.
In Fig. \ref{fig_gradient}, this may be the case for gradients $\Geff<0.3\unit{G/nm}$.

In a second set of experiments we investigate the spin relaxation rate as a function of rf field magnitude $\wone = \gamma B_1$.
Here we are able to measure $\itm$ for $\wone/2\pi$ between $60$ and $170\unit{kHz}$.
The rf field magnitude is calibrated by a spin nutation experiment \cite{poggio07}.
For small $\wone$, the spin correlation times are so short that
the measurement is dominated by the response time of the actively damped cantilever ($\sim 20\unit{ms}$).
The upper limit on $\wone$ is set by heating restrictions of the rf microwire.

\begin{figure}[t]
      \begin{center}
      \includegraphics[width=0.45\textwidth]{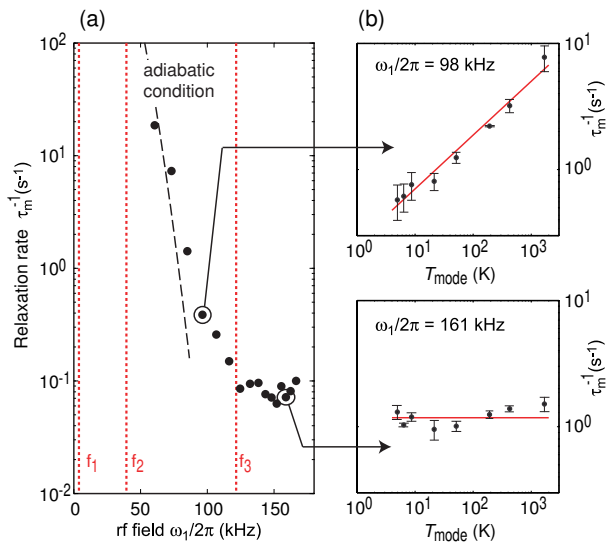}
      \end{center}
      \caption{(a) Spin relaxation rate $\itm$ as a function of rf field magnitude $\wone=\gamma B_1$, measured with setup {\bf B}.
      Each time $\wone$ falls below a cantilever mode, indicated by dotted vertical lines, a new dissipation channel is added.
      At low fields $\tm$ is presumably limited by the adiabatic condition (dashed line, Ref. \cite{adiabaticcondition}).
      (b) Spin relaxation rate $\itm$ as a function of mode temperature $\Tmode$, here for the third cantilever mode.
      The thermal motion of the mode is found to affect spin relaxation when $\wone/2\pi<f_3$ (upper figure),
      but not when $\wone/2\pi>f_3$ (lower figure). Solid line is a best fit.}
      \label{fig_driving}
\end{figure}

\begin{figure}[t]
      \begin{center}
      \includegraphics[width=0.45\textwidth]{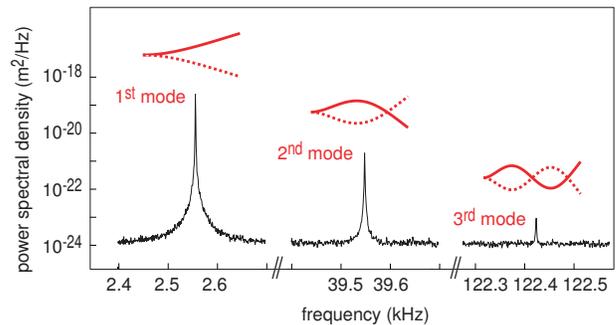}
      \end{center}
      \caption{Power spectral density of the thermomechanical noise at $T=4.5\unit{K}$, showing the first three vibrational modes
      of cantilever {\bf B}. Sketches illustrate the bending of the cantilever beam.
      Frequency, spring constant and quality factor of the lowest mode are $f_{1}=2.57\unit{kHz}$, $k_{1}=86\unit{\mu N/m}$ and $Q_{1}=17,000$.
      }
      \label{fig_modes}
\end{figure}

The measured relaxation rates are shown in Fig. \ref{fig_driving}(a). 
We find that the spin relaxation rate decreases rapidly until reaching a distinct knee at $\wone/2\pi\approx125\unit{kHz}$.
We believe that this feature is directly related to the thermal vibration of the third cantilever mode,
having a resonance frequency of $f_3=122\unit{kHz}$ [see Fig. \ref{fig_modes}].

To further explore the role of the third mode we can modify its effective temperature, denoted by $\Tmode$.
For that purpose we excite the mode with bandwidth-limited noise using a piezoelectric actuator mechanically coupled to the cantilever.
At the same time as we measure $\tm$, we also monitor the mean-squared motion $\xrms^2$ of the cantilever tip in that mode.
We can assign an equivalent mode temperature $\Tmode=\kci\xrms^2/\kB$, where $\kci$ is the effective spring constant of mode $i$ and $\kB$ is Boltzmann's constant.
Note that while $\Tmode$ can become very large for strong mechanical actuation, the actual (or bath) temperature of the cantilever
remains at $T=4.5\unit{K}$, as the heat capacity of the single mode is much smaller than
the heat capacity of the cantilever's phonon bath and the modes are only very weakly coupled.

We investigate the dependence of $\itm$ on $\Tmode$ at several different $B_1$ fields and find
the two different behaviors shown in Fig. \ref{fig_driving}(b):
If $\wone/2\pi$ lies above $f_3$, no change in spin relaxation is observed even for strong actuation.
On the other hand, for $\wone/2\pi$ below $f_3$, relaxation is greatly enhanced.
We can fit the dependence of the relaxation rate on the mode temperature and find $\itm \propto (\Tmode)^{0.43\pm0.07}$.
Because the slope does not level off even for low $\Tmode$, the spin relaxation rate is set by the thermal fluctuations of the third cantilever mode
over the entire investigated mode temperature range, in particular at thermal equilibrium where $\Tmode = T$.
At $\wone/2\pi=98\unit{kHz}$, we can also try to enhance spin relaxation by actuating the second mode.
No influence is observed, as expected, because $\wone/2\pi>f_2$.

In order to better understand the dependence of the spin correlation time on field gradient, rf field magnitude, and temperature, it is worthwhile to connect $\tm$ to the cantilever's mode spectrum. We calculate the transition rate following the analysis of Mozyrsky \etal \cite{mozyrsky03}.
To obtain an expression for $\tm$ similar to Eq. (\ref{eq_tr}),
taking into account the time-dependence of the rf field frequency $\wrf(t)$ during cyclic spin inversion,
we assume that $\tm$ can be described as the average relaxation rate over a frequency sweep,
\begin{equation}
\itm = \frac{1}{T_c} \int_{0}^{T_c/2} dt\, \frac{\wone^2}{\weff^2(t)}\, \gamma^2 S_{Bz}(\weff(t)).
\label{eq_itm_int}
\end{equation}
(See also Ref. \cite{mozyrsky03}, Eq. (7)). Here, $T_c=1/f_1$ is the oscillation period of the fundamental cantilever mode,
$\weff(t) = \{[\wrf(t)-\wo]^2+\wone^2\}^{1/2}$ is the effective magnetic field \cite{slichter_adiabaticsweeps},
and $\wrf(t)$ is modulated from $\wo-\Dw$ to $\wo+\Dw$. Note that because $\Dw\gg\wone$,
the Rabi frequency $\weff$ traverses a broad range of frequencies during the sweep.
A key difference between $\tm$ and $\Tr$ is therefore that $\tm$ is sensitive to noise
in a frequency band set by $\wone\leq\weff(t) \underset{\sim}{<} \Dw$, while $\Tr$ is influenced by noise in the vicinity of $\wone$ only.

The magnetic noise spectrum $S_{Bz}(\omega)$ is given by
\begin{equation}
S_{Bz}(\omega) = \Geff^2 \sum_{i=1}^n
\frac{\kT}{\pi \kci \fci \Qci} \ \frac{(2\pi\fci)^4}{((2\pi\fci)^2-\omega^2)^2+(2\pi\fci \omega/\Qci)^2},
\label{eq_modes}
\end{equation}
where $n$ is the number of modes with a significant noise contribution. Because of the high quality factors $Q_i$, $S_{Bz}(\omega)$ exhibits
a discrete set of sharp peaks at the mode frequencies $f_i$. We can facilitate the analysis of 
Eq. (\ref{eq_itm_int}) by treating the peaks as $\delta$-functions, and find that
\begin{equation}
\label{eq_itm}
\itm \approx \frac{\gamma^2\Geff^2\kT}{2\Dw} \sum_{i=n'}^{n} \frac{\wone^2}{2\pi\fci\kci \sqrt{(2\pi\fci)^2-\wone^2}} ,
\end{equation}
where $n'$ is the lowest mode whose resonance frequency $f_{n'}$ is above $\wone/2\pi$.
In other words, when $\wone$ is reduced a new relaxation channel is added each time
$\wone/2\pi$ becomes less than a cantilever resonance $\fci$. While $n$ can be large,
the lowest mode $n'$ will dominate (\ref{eq_itm}) because it has the lowest effective spring constant
and hence the largest thermal vibration amplitude.

As a result, we find that Eq. (\ref{eq_itm}) describes both the kink in Fig. \ref{fig_driving}(a) at the frequency of the third mode and
produces the correct trends in $\Geff$ and $\Tmode$. Eq. (\ref{eq_itm}) and Refs. \cite{sidles92,mozyrsky03}, however, predict
that $\itm \propto \Geff^2\Tmode$ --- a significantly stronger dependence on $\Geff$ and $\Tmode$ than our experimental observation,
$\itm \propto \Geff^{1.23\pm0.16}\Tmode^{0.43\pm0.07}$.
We are aware of two possible reasons for this discrepancy: First, the analysis of the spin transition rate [Eq. (\ref{eq_itm_int})]
is based on the Bloch-Redfield approximation, where the magnetic noise is assumed to be uncorrelated \cite{slichter_blochredfield}.
This assumption may be violated because the correlation time of the noise, $\tau_{c,i} = \Qci/\pi\fci$, can be significant, possibly even longer than $\tm$.
Furthermore, we assume that spins are non-interacting, which will not necessarily be valid, especially at low rf fields where local fields can easily exceed $B_1$.

In conclusion, we find that nuclear spin relaxation can be induced by a single, low-frequency mode of a micromechanical resonator.
The spin relaxation rate is observed to increase both with the gradient and the effective mode temperature.
We also find that long spin lifetimes are recovered when increasing the magnitude of the rf field to raise the Rabi frequency
above the lowest mechanical mode frequencies.

We thank C. Rettner, M. Hart and M. Farinelli for fabrication of microwire and magnetic tip,
B. W. Chui for cantilever fabrication, D. Pearson and
B. Melior for technical support, and M. Ernst for helpful discussions.
We acknowledge support from the DARPA QUIST
program administered through the US Army Research Office,
and the Stanford-IBM Center for Probing the Nanoscale, a NSF Nanoscale
Science and Engineering Center. C. L. D. acknowledges funding from
the Swiss National Science Foundation.


\begin{thebibliography}{}
\bibitem{Geff}
The introduction of an effective gradient $\Geff$ is necessary because not all spins in the
resonant slice experience the same $\dxBzu$. We believe that an appropriate definition for $\Geff$ is
the weighted-rms value
$\Geff(\Bext) = \left\{ \sum_{k=1}^{n}  \left[\dxBzu({\bm r}_k)\right]^4 / \sum_{k=1}^{n} \left[\dxBzu({\bm r}_k)\right]^2 \right\}^{\frac{1}{2}}$,
because it gives more weight to spins contributing large magnetic forces $F_k^2 = \mu^2 \left[\dxBzu({\bm r}_k)\right]^2$ to the signal.
$\{\vecr_k\}_{k=1...n}$ are the positions of the spins for which $|\gamma\Btip(\vecr_k)-\wo|<\Dw$.
\end{thebibliography}
\end{document}